\documentclass[11pt]{article}

\usepackage[T1]{fontenc}
\usepackage[utf8]{inputenc}
\usepackage{mathpazo}

\usepackage[margin=1.25in]{geometry}
\usepackage{parskip}
\usepackage{hyperref}
\hypersetup{colorlinks=true, linkcolor=black, citecolor=black, urlcolor=blue}
\usepackage{enumitem}

\newenvironment{hangparas}[2]
  {\setlength{\parindent}{0pt}
   \setlength{\leftskip}{#1}
   \setlength{\parskip}{#2\baselineskip}
   \everypar={\hangindent=#1\hangafter=1}\ignorespaces}
  {\par}

\title{}
\author{}
\date{}

\begin{document}

\title{Against Totalitarianism\\[4pt] \large Introduction to \emph{Tractatus Quanticum}}
\author{Jenann Ismael and Huw Price}
\date{}

\maketitle

\begin{quote}
\noindent\textbf{Abstract:} \emph{Tractatus Quanticum} (Covoni and Rovelli 2026) is
described by its authors as `a re-editing, which takes quantum
mechanics into account, of Wittgenstein\textquotesingle s famous
\emph{Tractatus}.' The original \emph{Tractatus} appeared with an
introduction by Bertrand Russell. For \emph{Tractatus Quanticum,} that
role fell to us. This is the result.
\end{quote}

\emph{Tractatus Quanticum} is a short book, written in a form that makes
it deliberately difficult to introduce. Like its great predecessor --
Wittgenstein\textquotesingle s \emph{Tractatus Logico-Philosophicus}
(Wittgenstein 1922) -- it proceeds by numbered propositions. Each is
austere and compressed, each assumes the weight of its predecessors. The
book does not argue so much as show, persuade so much as orient. And,
like the original \emph{Tractatus,} it ends by suggesting that the
reader who has truly understood it will recognize, then, that its
propositions point beyond what any propositions can say. As the authors
themselves put it, they are shamelessly parodying a Master. But the
parody, if that is the right word, is serious. And the point of
departure from the Master, when it comes, is one of the most
consequential moves in the understanding of modern physics.

Wittgenstein's \emph{Tractatus} appeared with an introduction by
Bertrand Russell. For \emph{Tractatus Quanticum,} that daunting role
falls to us. We will attempt to do for Covoni and Rovelli what -- with
mixed results -- Russell attempted to do for Wittgenstein: to prepare
the reader for what is coming, to situate it in its intellectual
context, and to draw out what seems to us the central nerve of the
argument. Famously, however, Russell\textquotesingle s introduction was
disowned by Wittgenstein, who felt that Russell had attended too closely
to the logical machinery, and not enough to the spirit that animated it.

Writing to the publisher Ludwig von Ficker in 1919, Wittgenstein said
that his book consisted of two parts, the part that was written and the
part that was not written.\footnote{Wittgenstein (1979), 94.} The second
part was the important one, and Wittgenstein was to feel that Russell
hadn't understood it. Lest we repeat Russell's mistake, our observations
about the unwritten part of \emph{Tractatus Quanticum} will be careful,
and expressed in the same silent vocabulary. What follows here is merely
the residue -- our written comments on the written text. For these, we
begin where Wittgenstein began, with an audacious claim about the nature
of the world. Covoni and Rovelli begin there, too, though they
immediately revise what they find.

\section{Wittgenstein against totalities}

At the beginning of the \emph{Tractatus,} Wittgenstein doesn't mess
around. In the very first lines, he tells us what the world is:

\begin{quote}
1 The world is everything that is the case.

1.1 The world is the totality of facts, not of things.
\end{quote}

This was already a \emph{scharf} move. The world isn't a collection of
\emph{objects} sitting around awaiting description, but \emph{facts} --
states of affairs that obtain. In the propositions that followed,
Wittgenstein\textquotesingle s purpose was to show the logical structure
that any language adequate to describing such a world must possess --
and, more importantly, to reveal the limits of what such a language can
say.

Those limits turn out to be severe. Russell identified what he took to
be Wittgenstein\textquotesingle s fundamental thesis in his
introduction:

\begin{quote}
{[}I{]}t is impossible to say anything about the world as a whole, and
whatever can be said has to be about bounded portions of the world.
(Russell 1922, 15)
\end{quote}

The reason is not technical but structural. We could only say something
about the world as a whole if we could get outside it -- if, for a
moment, it ceased to be for us the whole world, and became instead an
object within a larger totality. We could then survey the world from
some Archimedean point, within this larger realm. But there is no such
larger totality. By definition, the world is everything. As Russell put
it, following Wittgenstein\textquotesingle s own analogy: our field of
vision has no visual boundary, not because it is infinite, but because
there is nothing outside it for a boundary to be a boundary against.

For Wittgenstein, the Self suffers the same fate. His twin metaphysical
targets -- the ``I'' that views the world as if from a point outside it,
and the world so viewed -- stand or fall together. Each is no more to be
found among the facts of the world than the world\textquotesingle s own
totality is. The Self is the limit of the world, not something within
it.

This is the double vanishing act at the heart of the \emph{Tractatus.}
World-as-ultimate-object and self-as-ultimate-subject are two sides of
the same illusory coin. Neither can be located within the world. When we
try to talk about them, we push language past its limits. The result is
nonsense -- in Wittgenstein's enduring metaphor, the rungs of the
ladder, which must be climbed and then kicked away.

What gets kicked away is not merely a list of unspeakable topics -- say,
God, ethics, the self, the meaning of life. What gets kicked away is the
entire framework of the book: the picture theory of meaning, the notion
of atomic facts, the say/show distinction itself. Proposition 6.54 is
clear: "Anyone who understands me eventually recognizes {[}these
propositions{]} as nonsensical."

\section{Kicking even more away}

\emph{Tractatus Quanticum} inherits all of this self-immolatory
ambition, and manages to to extend it. In the original \emph{Tractatus,}
there is no god\textquotesingle s-eye view outside of the world from
which one could take in the totality of existence, no position outside
the world from which its boundedness is visible and statable. What
quantum mechanics adds, Covoni and Rovelli argue, is that there is no
view at all, outside of a perspective. The big story of quantum
mechanics is that it brings a new logic of perspectives, a logic
different from the classical one, that does not allow the construction
of an absolute, non-nonrelational description of reality.

Each of us sees the world from a particular perspective, of course. But
the world of classical physics was a shared space, of objects and
intrinsic properties available in principle to any observer. Observers
looking at the same scene from different locations, or through different
sensory pathways, will have different experiences. But those differences
are explained by differences in their relations to the same objects, in
the classical view. It's a single world. Variations among the
experiences of observers are explained by differences in how they couple
to it, and those differences themselves can be described in absolute
terms.

According to Covoni and Rovelli, what goes missing in quantum mechanics
is this non- relational core from which perspectives can be
reconstructed. In quantum mechanics there are perspectives and relations
between perspectives, but the theory no longer permits the extraction by
triangulation of a public world of objects and properties that the
various perspectives are perspectives \emph{on}. It's perspectives all
the way down: perspectives on perspectives on perspectives.

There is a parable that Paul Benaceraff tells in a different
context.\footnote{One of us (JTI) heard the parable from Benacerraf
  personally. It concerned the foundations of mathematics -- Hilbert and
  philosophers panicking about the storm caused by Gödel, and the castle
  of mathematics standing strong.} The story goes that there were
spiders in the basement of a HUGE, beautiful castle, reaching into the
sky. The spiders were building a vast web. They worked tirelessly,
making it more and more elaborate and intricate. One day a big storm
came, and blew through the basement, destroying the web. The spiders all
panicked, because they thought that the web was holding up the castle.

Covoni and Rovelli take quantum mechanics to show that the absolute,
classical world was the webs in our intellectual basement -- like the
castle, the network of perspectives stands up without it. The same
reasoning that shows in classical physics that there is no god's-eye
view outside the world, from which the world appears as a limited whole,
also shows that in quantum mechanics there is no view outside of a
perspective -- no standpoint from which the world appears in
non-relational form.

Just as Wittgenstein says that there is nothing outside the world,
\emph{Tractatus Quanticum} argues that there is nothing outside of a
perspective -- no world that all perspectives are perspectives
\emph{on}. What remains on the absolute side of the ledger, once we have
factored out all of the perspectival elements, is an empty shell. The
common object of all of the different possible perspectives was always a
mirage.

That frees the relations between perspectives from the strict discipline
imposed by the logic of `looking at the same world from different
locations'. Whether a particle has a definite position, a definite
momentum, a definite spin -- these are not questions with
perspective-independent answers. The answer depends on what has
interacted with what. Facts are facts relative to a perspective, which
is to say relative to a physical system, which is to say relative to a
set of interactions and correlations. "All things stand only in relation
to a perspective," as Proposition 3.7 of \emph{Tractatus Quanticum} puts
it, "and with a shift of perspective, what once appeared as fact may be
absent or incompatible with other facts from another perspective."

The same move collapses the distinction between public and private. In
classical physics, objects and their properties were detached from
experience, and housed in the public landscape. Experiences were
relegated to a private inner world of experience. Objects were
characterized intrinsically by quantities like mass, charge, position,
momentum -- that they carried with them independently of any observer or
interaction. But we knew them through the patterns of light color, sound
and smell that they caused in us. In \emph{Tractatus Quanticum} that
separation of inner and outer, public and private, (literally)
disappears.

A perspective, in this framework, is not a mind. It is not a conscious
observer, a biological system, or anything specifically human. It is a
physical system, standing in relations of correlation with other
physical systems -- a thermometer, a detector, a brain, a particle. What
a system knows about another system is just the set of correlations
between them, physically embodied, in principle accessible to any
further system that interacts with the first. Perspectives are
transparent to each other because they are themselves physical facts,
visible from elsewhere. Knowing another perspective is simply a matter
of coupling to it. In this way, Covoni and Rovelli dissolve the apparent
mystery of how one perspective can know another, without reintroducing
the absolute standpoint.

The parallel with the original \emph{Tractatus} is now direct, and it
runs deeper than might first appear. In Wittgenstein, two things
vanished from the sayable together: the world as totality, and the self
as metaphysical subject. They vanished together because they were two
aspects of a single impossible standpoint -- the view from nowhere. The
world as totality requires a perspective outside the world, from which
its boundedness is visible; and the metaphysical subject is precisely
that perspective, the point from which a world is a world. Each implies
the other, and neither can be found among the facts. They are the paired
limits of the sayable, and when the ladder is kicked away, they go
together.

But the classical picture of the world preserved a stable middle ground,
between these two vanishing points. Classical physics gives us a public
landscape. Objects and their intrinsic properties -- mass, charge,
position, momentum -- are housed in a shared space, independent of any
observer, available in principle to all perspectives. The self as
metaphysical subject might be unsayable, and the world as totality might
be felt rather than stated, but the world of physics was neither of
these extremes. It was the common object of which all perspectives had
their own view. Differences between observers -- the yellow look of
things to one, the molecular wavelengths to another -- were explained by
differences in how each observer coupled to the same underlying reality.

This classical picture is reflected in the distinction between primary
and secondary qualities, drawn by natural philosophers in the
seventeenth century. As Galileo puts it in 1623:

\begin{quote}
{[}T{]}astes, odours, colours, etc., on the side of the object in which
they seem to exist, are nothing else than mere names, but hold their
residence solely in the sensitive body; so that if the animal were
removed, every such quality would be abolished and
annihilated.\footnote{\emph{Il Saggiotore,} from a passage quoted by E.
  A. Burtt (1954), p. 85. Similar thoughts had been voiced much earlier.
  Democritus is said to have held that reality is simply a matter of
  "atoms and the void," and that all else is "by convention." But Burtt
  emphasizes that Galileo\textquotesingle s version of the doctrine
  introduces a new distinction between the objective realm of concern to
  mathematical science, on the one hand, and the subjective realm of the
  secondary qualities, on the other: "\emph{In the course of translating
  this distinction of primary and secondary into terms suited to the new
  mathematical interpretation of nature, we have the first stage in the
  reading of man quite out of the real and primary realm"} (Burtt 1954,
  89, italics in the original).}
\end{quote}

In this conception, secondary qualities belong to the perceiver, primary
qualities to the object. And while the god's-eye view that surveyed the
totality remained mystical, the physicist's view -- the description of
objects as they are in themselves, stripped of perspective -- was the
working aspiration of science. Self and world might be the paired limits
of the sayable, but between them stood a solid public world -- of which
perspective-independent description was not only \emph{possible,} but
the whole point.

Covoni and Rovelli take quantum mechanics to abolish this middle ground.
What the theory reveals is that there are no intrinsic properties
waiting in the public landscape, no perspective-independent values for
position, momentum, or spin that objects carry with them prior to and
independently of interactions. A variable may simply have no value in a
given perspective. Facts are facts relative to a perspective --
relative, that is, to a physical system standing in relations of
correlation with other physical systems. And this means the classical
picture's organizing assumption collapses: there is no common object
that all perspectives are perspectives on, no shared world behind the
window that each observer looks through from their particular angle.

The private inner world of experience, which in the classical picture
was the subjective counterpart to the objective public landscape, is
also gone. The distinction between inner and outer, between the private
world of the perceiver and the public world of physics, depended on the
same assumption -- that there was a perspective-independent reality to
be on one side of. Quantum mechanics removes the assumption, and both
sides of the distinction go with it.

With characteristic candor, Russell concluded his introduction to the
\emph{Tractatus} by noting that he was not sure whether
Wittgenstein\textquotesingle s central thesis was ultimately true.
Russell found himself unable to accept that nothing could be said about
the world as a whole. He suspected there must be some way to speak about
the totality of things, even if logic made it difficult. For his part,
Wittgenstein felt that Russell had missed the point entirely -- not the
logical point but the human one. The book was not primarily about the
limits of formal language. It was about the limits of everything --
thought, world, self -- and what those limits, rightly felt, opened
onto: the mystical, which was neither a doctrine nor a silence, but a
way of being oriented toward existence.

\emph{Tractatus Quanticum} is not a mystical book. Its authors are
physicists and philosophers, and they write with the confidence of
people who believe that quantum mechanics has settled something that
Wittgenstein could only approach obliquely. But they share with
Wittgenstein the conviction that good philosophy dismantles idols rather
than erecting them, and that the most important thing philosophy can do
is show us, over and over again, that certain questions have no meaning
-- not because the answers are hard to find, but because the questions
themselves rest on assumptions that the structure of reality does not
support.

The assumption Covoni and Rovelli are after is the oldest one in
philosophy: that there is a way things are, absolutely, from no
perspective, available in principle to any mind sufficiently clear and
powerful. The \emph{Tractatus} denied this about the world as a whole.
\emph{Tractatus Quanticum} denies it about everything. There is no world
behind the perspectives, stocked with intrinsic properties, waiting for
the right observer to arrive and find it as it really is. There is no
self behind the perspectives either, pure and transparent, surveying the
scene from a standpoint that belongs to no particular location in the
physical order. There are perspectives -- physical, partial, relative,
transparent to each other because they are facts -- all the way down.

\section{Pluralism against totalitarianism}

Finally, to a third reason why there can be no single world. It is tied
to a major shift in Wittgenstein's conception of language, between the
\emph{Tractatus} and his later work. As we imagine that Covoni and
Rovelli will agree, quantising the lessons of the \emph{Tractatus}
should not blind us to Wittgenstein's own mature view of its failings --
and these offer a different challenge to the idea of the world as a
single totality.

In the \emph{Tractatus,} Wittgenstein's view of language is the one
summarised by Russell in his Introduction: ``The essential business of
language is to assert or deny facts.'' (Russell, 8) Russell isn't
challenging this view, or the view of the world that Wittgenstein takes
to go with it. On the contrary:

\begin{quote}
The world consists of facts: facts cannot strictly speaking be defined,
but we can explain what we mean by saying that facts are what make
propositions true, or false. (Russell 1922, 10)
\end{quote}

Famously, however, Wittgenstein came to reject this view about ``the
essential business of language.'' In his \emph{Philosophical
Investigations} (PI, Wittgenstein 1953), one of his main themes is that
we do many things with language.

\begin{quote}
We remain unconscious of the prodigious diversity of all the everyday
language-games because the clothing of our language makes everything
alike. (PI, 224)

Think of the tools in a tool-box: there is a hammer, pliers, a saw, a
screw-driver, a rule, a glue-pot, glue, nails and screws.---The
functions of words are as diverse as the functions of these objects.
(And in both cases there are similarities.) Of course, what confuses us
is the uniform appearance of words when we hear them spoken or meet them
in script and print. For their \emph{application} is not presented to us
so clearly. Especially when we are doing philosophy. (PI, \#11)
\end{quote}

In the very next paragraph, Wittgenstein gives us another metaphor for
this view of language -- multiple functions, obscured by superficial
similarities.

\begin{quote}
It is like looking into the cabin of a locomotive. We see handles all
looking more or less alike. (Naturally, since they are all supposed to
be handled.) But one is the handle of a crank which can be moved
continuously (it regulates the opening of a valve); another is the
handle of a switch, which has only two effective positions, it is either
off or on; a third is the handle of a brake-lever, the harder one pulls
on it, the harder it brakes; a fourth, the handle of a pump: it has an
effect only so long as it is moved to and fro. (PI, \#12)
\end{quote}

Not all of \emph{Tractatus Quanticum}'s readers will share
Wittgenstein's evident familiarity with the cabin of a steam locomotive,
but the idea is clear. Tools that look similar can do different jobs.

Let's extend Wittgenstein\textquotesingle s metaphor a little. In
addition to all the \emph{handles} in the cabin of a locomotive, there
are other kinds of control devices, also adapted to the capacities of
human operators. There may be foot pedals and finger buttons, for
example -- perhaps touch screens, in particularly modern steam
locomotives. With this extension, we can use the distinction between
handles and other kinds of controls as an analogy for a distinction
between language games. Some of the things we do with language can be
done by \emph{making a statement --} by saying \emph{that} something is
the case, ``It is going to rain'', for example, or ``The president is a
moron''. But other things we do with language rely on different kinds of
utterances. ``Is it going to rain?'' is a question, not a statement.
``Don't vote for the moron!'' is a request, or perhaps an order, not a
statement. Let's think of statements as handles, and the other kinds of
language, like questions and commands, as other kinds of controls --
everything that isn't a handle.

As a test for whether an utterance falls on one side or other, we can
ask whether it makes sense to add ``It is the case that \ldots'' at the
beginning. We can do this with statements, and it doesn't seem to change
the meaning. ``It is the case that the president is a moron'' is
perfectly grammatical, and it says much the same as the original. (``The
president is a moron''.) But ``It is the case that is it going to rain''
makes no sense at all. (Read it carefully.) Is it a question? Your guess
is as good as ours!

This distinction is rough in various ways. For one thing, there are many
things we can do with language on either side of the line. Compare
``Where's the money?'' and ``I'm asking you where the money is'', for
example. Something similar is true in our updated version of
Wittgenstein's locomotive. Some of the jobs that are done by handles
could also be done by other control devices, and vice versa. If steam
locomotives had ejector seats, the control could be a handle, a button,
or a foot pedal. (A touch screen might be too slow.)

This means that even if we stick to the statement-making uses of
language -- to sayings \emph{that} something is the case -- we are still
dealing with lots of late-Wittgensteinian diversity. Just as there are
many things we can do with \emph{handles} -- and surely no totality of
all the possible things we \emph{could} do with handles -- so with
statements.

Here's an example. One thing we do language is to issue warnings, such
as ``Be careful of tigers!'' This doesn't tell our audience that
something is this case -- we can't say ``It is the case that \emph{be
careful of tigers}.'' But it is easy to move the act of giving a verbal
warning to the other side of the line. One way to do it, which works for
many kinds of advice and instructions, is just to say ``You should be
careful of tigers.'' (It is easy to add ``It is the case that \ldots''
to that -- you should try it and see.) Another way is to invent a label
for things we should be careful about, as when we say ``Tigers are
dangerous.''

Why does this matter? Because where go statements, there go facts, too,
at least in an everyday sense. ``It is the case that P'', ``It is true
that P'', and ``It is a fact that P'' are pretty much interchangeable,
for any sentence P that fits the general grammatical format. This means
that once we start warning our neighbours about tigers by saying
``Tigers are dangerous'', we've got \emph{facts} about danger.

What's the big deal, you may think. If ``Tigers are dangerous'' isn't a
fact, what is? But remember where we started, with Wittgenstein's and
Russell's agreement that the ``world consists of facts'' (Russell 1922,
10); that ``{[}t{]}he world is everything that is the case''
(\emph{Tractatus,} Proposition 1). Taken literally, this means one of
two things. Either we \emph{added} something to the world when we
invented the term \emph{dangerous,} as a new way of warning folk to be
careful -- we added facts about danger. Or these facts about danger were
\emph{already out there,} as part of the world, waiting for us to invent
a word to talk about them.

Neither option sits very comfortably with the idea that the world is a
totality, fixed in advance, independently of what we humans happen to do
with language. The first option is flatly incompatible with this idea,
because it holds that we added something new to the world, when we
started to talk about danger. The second avoids this problem, but only
by populating the world with endless kinds of facts, for most of which
we never invent the words.

In different ways, then, Wittgenstein's open-ended conception of the
range of possible language games leads to problems for totalitarianism
-- the conception of the world as a totality of facts. There are simply
too many kinds of facts.

In response, we might try to distinguish the ``genuine'' facts, so that
we could go on saying that the world is everything that is the case,
\emph{in this distinguished sense.} Many thinkers, physicists among
them, have been attracted to the idea that the distinguished facts are
those of physics. In the 1920s, this idea was popular among the Vienna
Circle, themselves influenced by the \emph{Tractatus}. It echoes the
seventeenth century distinction between the primary qualities -- those
of physics -- and everything else.

As we have seen, \emph{Tractatus Quanticum} raises new challenges for
this idea. It argues that \emph{physics itself} undermines the idea of a
single, objective, perspective-independent world. But there were prior
challenges from other quarters. Probability seems a fundamental notion
in physics, for example. Yet a long tradition maintains that probability
simply cannot be understood, without highlighting its link to the
viewpoint of agents acting under uncertainty. That threatens to put
another kind of perspective at the heart of the supposedly distinguished
realm -- probability is central in quantum mechanics, for example.

One way to express this point is to say that probability is a secondary
quality, in a modern extension of the seventeenth century
classification. The threat is then that physics cannot stay on the
primary side of the line, as Galileo and his intellectual descendants
intended. Some philosophers have gone even further. The American
philosopher Hilary Putnam -- himself, like Wittgenstein, a late but
influential convert to pragmatism -- suggested that Kant held that
everything was a secondary quality, in effect.\footnote{Putnam (1981),
  60--61.}

By this path, some philosophers have reached conclusions that seem
similar to those of \emph{Tractatus Quanticum.}\footnote{\emph{Tractatus
  Quanticum} itself, by a different path, thus arrives at the same place
  as Dewey: "If we see that knowing is not the act of an outside
  spectator but of a participator inside the natural and social scene,
  then knowledge resides in the consequences of directed action.''
  (Dewey 1929, 27)} The audacity of \emph{Tractatus Quanticum} is that
it does so from within physics. If Covoni and Rovelli are right, then
with quantum theory Galileo's goose lays its final egg. This egg has
three things in common with the mythical golden one: (i) it is a
treasure, in this case by philosophical standards; (ii) it occasions the
demise of its own mother; and (iii) it is the end of the line, in the
sense that it contains no further poultry of the same Galilean variety.

\section*{Bibliography}
\begin{hangparas}{1.5em}{1}

Burtt, E. A. 1954. \emph{The Metaphysical Foundations of Modern Physical
Science.} Garden City, N.Y.: Doubleday Anchor Books.

Covoni, N. \& Rovelli, C. 2026. \emph{Tractatus Quanticum.}
\url{https://arxiv.org/abs/2512.06034}

Dewey, J. 1929. \emph{The Quest for Certainty: A Study of the Relation
of Knowledge and Action}. Minton, Balch \& Company.

Putnam, H. 1981. \emph{Reason, Truth and History.} Cambridge: Cambridge
University Press.

Russell, B. 1922. `Introduction to Ludwig Wittgenstein's
\emph{Tractatus Logico-Philosophicus}', in Wittgenstein 1922, 7--19.

Wittgenstein, L. 1922. \emph{Tractatus Logico-Philosophicus}. C. K.
Ogden (trans.), London: Routledge \& Kegan Paul. Originally published as
``Logisch-Philosophische Abhandlung'', in \emph{Annalen der
Naturphilosophische}, XIV (3/4), 1921.

Wittgenstein, L. 1953. \emph{Philosophical Investigations}. G.E.M.
Anscombe and R. Rhees (eds.), G.E.M. Anscombe (trans.), Oxford:
Blackwell.

Wittgenstein, L. 1979. `Letters to Ludwig von Ficker'. In C. Grant
Luckhard (ed), \emph{Wittgenstein, Sources and Perspectives.} 1979.
Ithaca, N.Y. : Cornell University Press, 82--98.

\end{hangparas}

\end{document}